\documentclass[12pt,letter]{article}
\usepackage[english]{babel}
\usepackage[utf8]{inputenc}
\usepackage{hyperref}
\usepackage{graphicx}
\usepackage{geometry}
\title{Pi\`{e}ces de viole des Cinq Livres and their statistical signatures: the musical work of Marin Marais and Jordi Savall}

\author{Igor Lugo$^1$\thanks{Corresponding author: \href{igorlugo@crim.unam.mx}{igorlugo@crim.unam.mx}}, Martha G. Alatriste-Contreras$^2$}

\date{%
    $^1$Centro Regional de Investigaciones Multidisciplinarias, Universidad Nacional Aut\'{o}noma de M\'{e}xico, Cuernavaca, Morelos, M\'{e}xico\\%
    $^2$Facultad de Econom\'{i}a, Universidad Nacional Aut\'{o}noma de M\'{e}xico, Ciudad de M\'{e}xico, M\'{e}xico\\[2ex]%
    \today
}

\begin{document}
\maketitle

\begin{abstract}
This study analyzes the spectrum of audio signals related to the work of ``Pi\`{e}ces de viole des Cinq Livres'' based on the collaborative work between Marin Marais and Jordi Savall for the underlying musical information. In particular, we explore the identification of possible statistical signatures related to this musical work. Based on the complex systems approach, we compute the spectrum of audio signals, analyze and identify their best-fit statistical distributions, and plot their relative frequencies using the scientific pitch notation. Findings suggest that the collection of frequency components related to the spectrum of each of the books that form this audio work show highly skewed and associated statistical distributions. Therefore, the most frequent statistical distribution that best describes the collection of these audio data and may be associated with a singular statistical signature is the exponential.\\
\\
keyword: Bass viol de Gamba, Marin Marais, Jordi Savall, complex systems, statistical distributions
\end{abstract}

\section*{Introduction}
Marin Marais is one of the most outstanding composer and performer musician of bass viola da gamba in music history. Not only in his era, but also in current days, his music has continued to delight from novice to expert musicians. Nowadays, Jordi Savall---a contemporary conductor, composer, historian, and viol player---has communicated globally most of the Marias' work. The extraordinary musical contribution of both musicians has showed one of the highest level of musical expression over time. There is a deep musical connection between this pair of musicians (named in the following as Marais-Savall) that has transcended time and frontiers. However, a scientific analysis of Marais-Savall audio signals---waveforms---is still missing in the musician and scientific communities. In particular, one of the most important works of Marais, \href{https://www.alia-vox.com/en/catalogue/marin-marais-pieces-de-viole-des-cinq-livres/}{``Pi\`{e}ces de viole des Cinq Livres,''} has not been explored based on its statistical properties that underly its music information. Therefore, our study aims to analyze the audio signals of these five volumes for identifying the statistical distributions that best describe their spectrum---the frequency of components related to music notes. After establishing the presence of such distributions, we can reinterpret the music of bass viol and identify the signature of viol players.

This statistical approach for using waveforms and its spectrum has been applied by Lugo and Alatriste-Contreras~\cite{LugoAlatriste2022}. They suggested that the concept of virtuosity in music is possible related to entropy values and the best-fit distributions of the spectrum of audio signals. In particular, they suggested that the waveform and its spectrum contain information to identify levels of virtuosity in music. Moreover, the work of Downey~\cite{Downey2016} showed different topics of signal processing in music. In particular, he presented techniques and applications with a programming-based approach for understanding real audio signals. Other relevant work of audio signals is M\"{u}ller and Klapuri~\cite{MullerKlapuri2014}. They presented an overview of principles and applications of music signals that are the key for underlying music analysis problems. Therefore, the data analysis of audio signals based on interdisciplinary approaches and the current digital technology may generate a deep understanding of the underlying information in music. The intuition of musicians about identifying a particular composer when only playing or hearing few notes of some music repertoire can be confirmed if we look into the statistics of the spectrum of signals.

In the case of Marais-Savall, the identification of their unique statistical signatures might be related to different aspects of their lifes. About Marais, several studies about his musical abilities have highlighted factors that are associated with his personal experience and social relationships~\cite{Cyr2016, Bane2018}. In particular, the work of Milliot and de la Gorce~\cite{MilliotGorce1991} described almost a complete view of the context in which Marais' developed his musical creativity and skills. For example, his relationship with two of the most respected musicians in that time, Jean de Sainte-Colombe and Jean-Baptiste Lully. Other work that complemented this reference is the audio work of Savall et al.,~\cite{Savalletal2010}. It offers materials that not only can listened, but also read; we can read information about the collection of the audio tracks. For example, the booklet described the common discussion about musicians in that time about the balance between melody and harmony. Finally, an unexpected, but interesting study is the work of Matloubieh et al.,~\cite{Matloubiehetal2020}. This study came from another discipline, and authors suggested that Marais' mixed his musical reputation with a medical procedure about lithotomy. Then, the influence of Marais' covered not only common issues on music, but also different and relevant themes on his time. 

In the case of Savall, there are different information sources that display his work across several areas---i.e., concert performer, teacher, researcher, just to name a few~\cite{FernandezTamaro2022}. For example, the websites of \href{https://www.alia-vox.com/en/artists/jordi-savall/}{AliaVox} and \href{https://www.fundaciocima.org/la-fundacio/?lang=en#1484565170801-dda387ae-eea4}{Fundaci\'{o} Centre International de M\'{u}sica Antiga} illustrate his outstanding work for rescuing and preserving early music. The work of Forti i Murrugat~\cite{FortiMarrugat2019} analyzed Savall's projects related to this type of music to propose a musical framework for music and art. Therefore, the Marais-Savall relationship shows singular musical and personal characteristics that are possibly imprinted in most of their musical work.

Our main questions are the following: does the collection of the five books show similar statistical distributions? and, what are those statistical distributions? We believe that the underlying attributes of waveforms and their audio spectrum are related to statistical distributions. They can be interpreted as the signature of a set of music repertoire. Depending on the performance of musicians---different musicians play the same repertoire---the signature varies only slightly from its real value. Therefore, a large set of audio signals of bass viol played by the same musician provides the event-based condition for identifying accurately the signature of a particular musical work.    

The document is divided on four sections. The Materials section shows the audio resources from the data that was collected. The Method section explains the application of the complex systems approach into an explorative data analysis based on identifying statistical distributions. The Results section displays our findings. Finally, the Discussion section points out some items to be considered in the analysis and gives the conclusions. 

\section*{Materials}
The data consisted of the audio material related to the work of Savall et al.~\cite{Savalletal2010}. This material, named~\href{https://www.alia-vox.com/es/catalogo/marin-marais-pieces-de-viole-des-cinq-livres/}{Pi\`{e}ces de viole des Cinq Livres}, is a collection of five audio CDs that contains a total of 84 tracks. We selected this audio material because is one of the best recording audio data about Marin Marais up until now. This material combines together audio recordings and historical documents that more accurately described the Marais' musical contribution. As a listener, the execution behind each track reflects Savall's expertise of playing the instrument and his knowledge of recording music. Because of the copyrights of this audio material, we suggest to obtain these CDs and follow our method for replicating results. Therefore, we transformed each track of this album from $m4a$ to $WAV$ files in 16-bit PCM. During this process, we spliced stereo to mono using~\href{https://www.audacityteam.org/}{Audacity}. 

On the other hand, we used different Python libraries to retrieve, analyze, and plot the data and results. In particular, we used the \href{https://numpy.org/}{Numpy}, \href{https://matplotlib.org/}{Matplotlib}, \href{https://www.scipy.org/}{Scipy}, and \href{https://pandas.pydata.org/}{Pandas}. Moreover, we used some part of the code provided by Downey~\cite{Downey2016}. The database and the code are available in our Open Science Framework (OSF) for the reproduction of our findings: \href{https://osf.io/2byqv/?view_only=34cf5c9b0fe442c4a7be6eb82f5fc203}{Complex systems and early music}.

\section*{Methods}
High quality audio recordings of viola da gamba are rare because it is not common to play such an instrument nowadays. In this case, Savall has provided a unique collection of recordings of Marais that we can analyze based on their audio signals. Therefore, in this section, we present our procedure for identifying the statistical signatures of the collaborative work between Marais-Savall.

The first step in this procedure is the use of the spectral decomposition. This is a procedure for simplifying audio data based on the Fast Fourier Transformation (FFT) algorithm and the Discrete Fourier Transformation (DFT)~\cite{Cooleyetal1965, Press2007,SciPy2021}. The result of this decomposition is named the ``spectrum'' that shows approximately the frequency components related to pitches---the dominant pitch and its harmonics. The importance of this spectrum is to identify the frequency components of sections related to the immediate musical composition or improvisation. The analysis of these sections is not trivial because they are commonly related to different musical interpretations in which the musician communicates emotions by performing. For example, contemporary guitarists improvise solos that engage  audiences or listeners to experience different emotions~\cite{Swarbricketal2019}. Then, the spectrum is one of the keys to unfold the rich musical imaginations of celebrated composer-performer musicians~\cite{Britannica2022}. Therefore, we used the spectral decomposition associated with the FFT as following:

\begin{eqnarray}
\label{eq:1}
	y[k] = \sum_{n=0}^{N-1} e^{-2\pi j \frac{kn}{N}} x[n]
\end{eqnarray}
in which $y[k]$ is the frequency component of a sequence of the signal $x[n]$ from $n$ to $N-1$. In our case, the identification of these components provides the inputs for understanding the statistical signature associated with our audio material. Therefore, we can define the statistical signature of a collection of audio materials as a clearly identified statistical distribution that shows particular properties.

Next, the second step is to analyze those collections of frequency components looking for the identification of statistical distributions that best describe them. The statistical distribution or probability distribution is a mathematical function that describes the occurrence of possible events. It approximates the generating process of particular data. A major advantage of this function is to infer properties underlying it~\cite{Ross2012,Ross2020}. In particular, continuous distributions are commonly related to three types of parameters: location, scale, and shape~\cite{LeemisMcQueston2008}. The location parameter, which is associated with the first moment or mean, refers to the place where the most frequent value is observed along the x-axis in a frequency plot. The scale parameter, which is associated with the second moment or variance, refers to how spread out are the data with respect to the location along the x-axis. The shape parameter, which is associated with all higher moments such as the skewness and kurtosis, refers to the shape or geometry of the data. It is important to mention that depending on the skewed or non-skewed data the location and scale parameters can be related to different measures, for example for skewed data is commonly suggested to compute the median and the entropy~\cite{CoverThomas2006,Smaldino2013,Mohretal2022}. Therefore, based on this collection of statistical measures, we can describe and infer the behavior of any distribution. In our case, after obtaining the frequency components, we are in the possibility of identifying the statistical distribution that represents accurately the unique audio signal of the work of Marais-Savall.

To identify the possible statistical distributions that best describe the audio work of Marais-Savall, we used the Kolmogorov-Smirnov (KS) test to identify whether or not our audio data comes from a certain distribution~\cite{Massey1951}. In our case, this test compares our frequency components (empirical data) with a set of given distributions (theoretical distributions). We used the normal, log-normal, exponential, Pareto, Gilbrat, power law, and exponentiated Weibull as our theoretical distributions. These distributions represent an important set of continuous distributions in the literature that might represent non-skewed and skewed data, as well as different relationships between them~\cite{Lehoczky2015, CollegeWilliamMary2022}. In particular, these distributions are connected with other distributions based on their properties, for example the linear combination, coevolution, and products, just to mention a few~\cite{LeemisMcQueston2008}. Therefore, the KS test using those statistical distributions provides an accurate process to compute and identify distributions that best describe the audio work of Marais-Savall.

\begin{figure}[!ht]
\centering
\includegraphics[width=9.5cm]{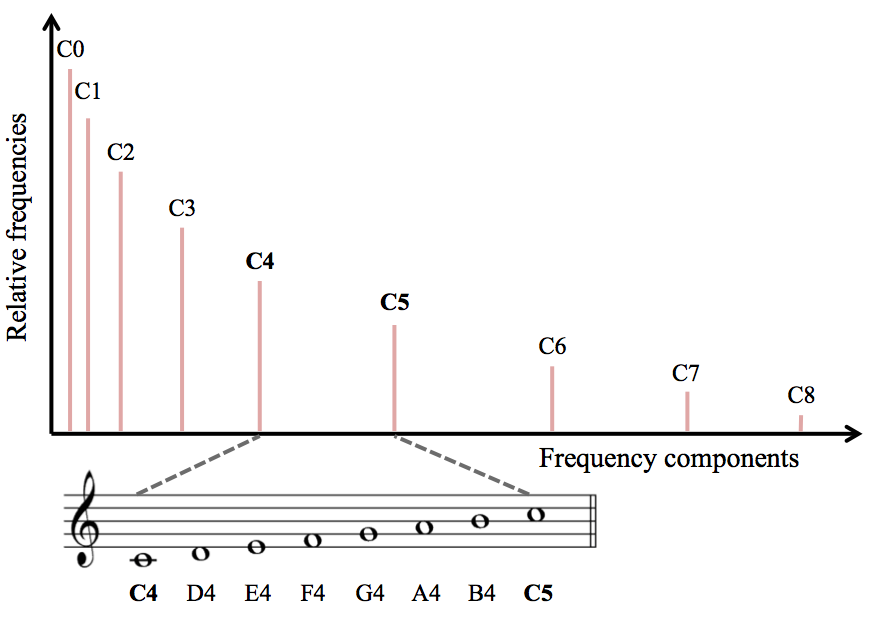}
\caption{{\bf Relative frequencies of the spectrum.} Relative frequencies are related to Hz, and Frequency components are associated with the output of the FFT (Eq.~\ref{eq:1}). We used the data provided by the Physics Department, Michigan Technological University~\cite{Suits2022}.
To avoid confusion in the scientific pitch notation, notes are translated by multiplying or dividing the frequency by 2. Then, in this Figure, vertical lines are approximations of octave locations used for reference purposes only.
}
\label{fig_sn}
\end{figure}

Finally, once the best-fit distributions were identified, we display their relative frequencies of the frequency components using the scientific pitch notation---notes names and octave numbers---as bins (Figure \ref{fig_sn}). This figure aims to show the statistical attributes of audio signals based on how frequently some notes and octaves are used. It is the key to understand the possible statistical signature of the work of Marais-Savall. In the next section, we display this figure in each result related to the book's audio data.

In essence, we use each track per book for computing the FFT. The resulted book's collection of component frequencies are analyzed for identifying its best-fit statistical distribution. To identify the distribution that best describes the component frequencies, we show our proposed plot of relative frequencies of the spectrum (Figure \ref{fig_sn}). Finally, the programming code of each step can be found in our project of the Open Science Framework platform.

\section*{Results}
The main goal of this study is to identify possible statistical signatures related to the spectrum of audio signals of the work of ``Pi\`{e}ces de viole des Cinq Livres'' based on the work of Marais-Savall. Each book, which is a collection of audio tracks, of this work was analyzed by following our proposed method. Therefore, for simplicity and ease of interpretation, we are going to show our results for the five books by using only our proposed plot related to the relative frequencies (Figure~\ref{fig_sn}).

Figure~\ref{tbestfit} shows the statistical signature of each of the books. As can be seen, the curves related to each of the collection of frequency components associated with their spectrums are highly skewed. This lack of symmetry suggested that lower and higher octaves are played more frequently. In particular, lower octaves represent the majority of notes played between the $C0$ and $C3$. On the other hand, higher octaves represent the relative notes played between $C6$ and $C8$. Moreover, between $C3$ and $C6$ octaves, we can see important differences of played notes. Books $4$ and $5$ represent the greatest difference, meanwhile books $1$, $2$, and $3$ are in between them. This particular result suggests that the main differences between the frequency components of each book are around the standard tuning $C4$.  

\begin{figure}[!ht]
\centering
\includegraphics[width=13cm]{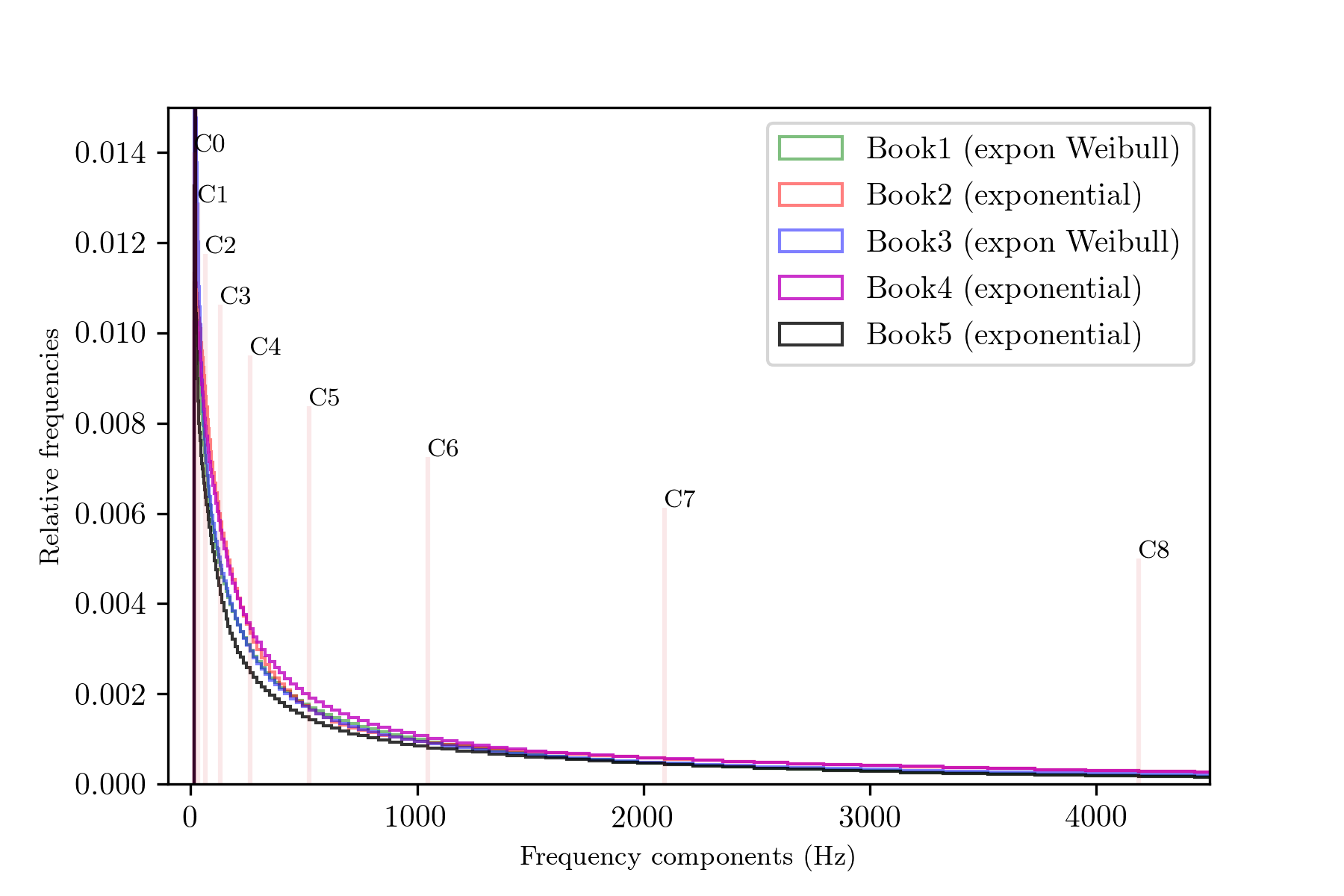}
\caption{{\bf Relative frequencies of the audio spectrums and best-fit distributions.} See Table~\ref{tbestfit} for statistical results of best-fit, parameters, and KS test results.
}
\label{5books}
\end{figure}

In addition to these results, we present the estimated parameters related to the location, scale, and shape per book. As we can see in the Table~\ref{btmoments}, there are two values related to the location: median and mean. In this particular case, we are interested in the median due to the resulted highly skewed distributions. By using the mean value, can be misleading or just plain wrong because it is commonly related to non-skewed data, such as the normal distribution. Then, the median values of the books associated with the exponential distribution show a stable location between the notes $A1$ and $E2$, meanwhile $book1$ and $book3$ showed locations less than $C0$. Next, there are two scale parameters related to the data dispersion: variance and entropy. As we have just mentioned, we used the entropy value due to its attributes related to skewed data. Then, entropy values showed similar dispersion except for the $book1$. Finally, the estimated shape parameters related to geometry showed that there is more weight on the right tail of the distributions, and they exhibit peaked shapes.

\begin{table}[!ht]
\centering
\caption{
{\bf Estimated parameters of frequency components per book}}
\begin{tabular}{|l|l|l|l|l|l|l|l|}
\hline
\multicolumn{1}{|c|}{\bf Name} & \multicolumn{1}{|c|}{\bf Median} & \multicolumn{1}{|c|}{\bf Mean} & \multicolumn{1}{|c|}{\bf Variance} &\multicolumn{1}{|c|}{\bf Entropy} &\multicolumn{1}{|c|}{\bf Skew} &\multicolumn{1}{|c|}{\bf Kurtosis}\\ 

\hline
$Book1$ & 11.9226 & 15.9130 & 165.1470 & 2.2294 &8.6444 &165.0793\\
$Book2$	& 79.3857 & 114.5291 & 13116.7683 & 5.7408 & 2.0 & 6.0\\
$Book3$	& 19.4983 & 58.5914 & 15103.0876 & 4.7947 & 7.5989& 126.6237\\
$Book4$	& 77.2398 & 111.4334 & 12417.3580& 5.7134 & 2.0& 6.0\\
$Book5$	& 54.4538 & 78.5601 & 6171.6320 & 5.3638 & 2.0 & 6.0\\
\hline

\end{tabular}
\begin{flushleft} 

\end{flushleft}
\label{btmoments}
\end{table}

Summing up, the results of our data analysis are conclusive for identifying reliable parameters that point out the statistical signature of the work of Marais-Savall. The frequency components of the audio spectrums related to each book were associated with highly skewed distributions. These distributions may well be related to the exponential distribution because it is the most frequent best-fit distribution presented in our results.

\section*{Discussion and conclusion}
Our study about the musical collaboration between Marais-Savall related to the audio work of ``Pi\`{e}ces de viole des Cinq Livres'' has showed the possibility of underlying its musical information. In particular, we could identify statistical attributes that distinguish the most frequent best-fit distributions related to each book's audio spectrums. Consequently, our results indicated that the frequency components of such spectrums must have been related to the presence of highly skewed distributions, particularly in relationship with the exponential statistical distribution. 

The significance of these findings is to be found in recognition of the musical work between musicians. Even though musicians are separated by time periods and places, their original and unique musical contributions can be recognized not only by the timbre, but also by the information related to the audio wave. In our case, the collaboration between Marin Marais and Jordi Savall has showed one of the highest levels of musical expressions over time that must have been recognized by their highly skewed distributions of the frequency components of audio spectrums. Such statistical distributions are related to the exponential distribution. This type of distribution is commonly related to describe system reliability and the times between events~\cite{Lehoczky2015}. One of its main characteristics is a constant failure rate function---no memory when considering events based on its age. In our case, the memoryless or Markovian~\cite{Billard2015} property indicates that if the most frequent octaves notes are played for $s$ units of times, the probability that higher octaves notes will play in additional time units is independent of $s$. In other worlds, the probability for playing lower or higher octave notes in an audio track is independent. Therefore, these findings suggest that the transition from one note to other in an audio track follows a random process based on the exponential distribution.

Translating this result into musical expressions, we can say that there may be a link between the improvisation and the selection of notes in particular musical passage. In the case of Marais-Savall, we know that the composition and performance abilities of Marais were frequently associated with improvisation~\cite{Savalletal2010,BritannicaME2022}. Consequently, it is expected that the Savall's interpretation and performance reproduce such Marais' habits. These findings may help us to understand that the free performance of the musician may follow different random processes that most of the time are related to skewed statistical distributions.

The implications of these findings regarding the teaching and learning of music are related to composing and playing activities that cover not only the bass viola da gamba, but also any type of string instruments. For example, in a musical composition, musicians may use the information related to type of statistical distribution for exploring alternatives or extensions to conceive a pice of music. Depending on the type of skewed distribution, musicians must use more frequently higher and lower octaves for achieving most of their musical material. Therefore, before starting the process of composition, it is recommended to analyze previous personal works and the work of other musicians for obtaining a unique and original material. In the case of performing, the prior information about statistical distributions can provide the keys for improvising music in different styles and contexts. For example, if we know that the work of Marais-Savall is best described by an exponential distribution, we must play the patterns suggested by such a distribution. We must play more frequently higher and lower octaves, meanwhile around the standard tuning, we can play notes for connecting and transitioning those octaves. Following this information, we can replicate the work of those musicians or generate our personal material.

Future studies on the current topic are therefore recommended. In particular, a natural extension of our findings is to explore the following questions: is it possible to find the presence of similar best-fit distributions whether a particular musical passage is played by different musicians? how different or similar can be the statistical properties of each performance? On the other hand, a future study with more focus on a computational approximation for composing music based on our findings is therefore suggested. To answer  these questions in future work, we suggested following the line of complex systems and music.
 
Therefore, the greatest contribution of this study is to underly statistical properties related to early music, composed and played by two outstanding musicians. Our method can be used for analyzing not only the bass viola da gamba, but also other string instruments and musicians. The audio work of ``Pi\`{e}ces de viole des Cinq Livres'' showed highly skewed distributions possible related to the exponential distribution. This type of statistical distribution may contain the keys to understand the elements of musical composition and performance of the bass viola da gamba.

\bibliographystyle{cmpj}
\bibliography{mybibfile01}

\section*{Supplementary material}
See Table~\ref{tbestfit} and~\ref{pdfs}. The criteria for selecting the final result in Table~\ref{tbestfit} were as follows: 
\begin{enumerate}
	\item Errors in the estimation method. Based on the \href{https://docs.scipy.org/doc/scipy/reference/generated/scipy.stats.rv_continuous.fit.html#scipy.stats.rv_continuous.fit}{scipy reference}, we used the Maximum Likelihood Estimation (MLE). If there were no errors, we selected the best fit; if there were errors, we selected the second best fit.
	\item Visualizing the Cumulative Distribution Function (CDF). If the estimated values of the KS test ($d$, p-value) of the first and second best fit were the same, we plotted the empirical and theoretical CDFs. This ensures to identify the best fit statistical distribution related to data.
\end{enumerate}
Consequently, we found that the fit of Pareto estimations showed RuntimeErrors. Then, we had to visualize the CDFs for selecting final results in Table~\ref{tbestfit}. For greater precision of the KS test estimated parameters and plot the CDFs, see and execute the code in \href{https://osf.io/2byqv/?view_only=34cf5c9b0fe442c4a7be6eb82f5fc203}{Complex systems and early music}. For the use of the same criteria applied to different scientific studies, see \cite{lugoAlatristeC2019}, \cite{lugoMartinezMekler2022}, and \cite{lugoAlatristeC2022}.

\begin{table}[!ht]
\centering
\caption{
{\bf Statistical attributes of the ``Pi\`{e}ces de viole des Cinq Livres''.}}
\begin{tabular}{|l|l|l|l|l|}
\hline
\multicolumn{1}{|c|}{\bf Name} & \multicolumn{1}{|c|}{\bf Best and second best fit} & \multicolumn{1}{|c|}{\bf Parameters} & \multicolumn{1}{|c|}{\bf KS test} \\ 
\multicolumn{1}{|c|}{} & \multicolumn{1}{|c|}{} & \multicolumn{1}{|c|}{\bf (a, b, loc, scale) } & \multicolumn{1}{|c|}{\bf (d, p-value)} \\ 

\hline
$Book1$ & Pareto  & (0.5000772790696841, & (8.810404621462098e-05, \\
	&  & -2.5313693644133393, & 0.7663125997309213) \\
	&  & 2.5320219554920884) &  \\
	& exponentiated Weibull *& (1.8078554913188745, &(8.810404621462098e-05, \\
	&  & 0.40464609484912933, & 0.7663125997309213) \\
	&  & 10.52751466016218) &  \\
\hline
$Book2$	& exponential* & (0.0007007909083517199, & (8.531262287569952e-05, \\
	&  & 114.52846083233646) & 0.785822040345603) \\
	& Pareto & (0.45694976715045765, & (8.531262287581054e-05, \\
	&  & -2.3484869507168717, & 0.7858220403442739) \\
	&  & 2.349187741621745) &  \\
\hline

$Book3$ & exponentiated Weibull * & (2.388182477333179, & (7.587900003558357e-05, \\
	&  & 0.40846064327782183, & 0.7629043858598745) \\
	&  & 0.0006176602639155107, &  \\
	&  & 8.883106480102926) &  \\
	& exponential & (0.0006176602639155108, & (7.587900003563908e-05, \\
	&  & 92.6733864816965) & 0.76290438585909) \\
\hline

$Book4$ & exponential *  & (0.00024278816105826503,  & (9.143090021945799e-05, \\
	&  & 111.43319977150223)  & 0.7129222750411135) \\
	& Pareto & (0.4717388601895959, & (9.143090021945799e-05, \\
	&  & -2.384176737481692, &  0.7129222750411135)\\
	&  & 2.3844195256376968) &  \\
	&  &  &  \\
\hline

$Book5$ & exponential* & (0.00036681097553665327, & (7.877884394258405e-05, \\
	&  & 78.55973558262349) & 0.8129868745072603) \\
	& Pareto & (0.4239939171603715, & (7.877884394258405e-05, \\
	&  & -1.6836235103637094, & 0.8129868745072603) \\
	&  & 1.683990321337741) &  \\
	&  &  &  \\
\hline

\end{tabular}
\begin{flushleft} Name of the books, the best and second best fit statistical distributions, the estimated parameters of the KS goodness-of-fit test, and the KS test two-sided statistic. * Statistics of the selected best fit test. See Table~\ref{pdfs} for the name of statistical distributions and their probability density function (PDF).
\end{flushleft}
\label{tbestfit}
\end{table}

\begin{table}[ht]
\centering
\caption{\label{pdfs}{\bf Statistical distributions and their probability density functions (PDF).} }
\begin{tabular}{|c|c|}
\hline
{\bf Name} & {\bf PDF}  \\
\hline
exponential	&  $f(x) = exp(-x)$, for $x >= 0$\\
Pareto		& $f(x, b) = \frac{b}{x^{b+1}}$, for $x>=1$, $b>0$ \\
exponentiated Weibull	& $f(x, \alpha, c) = \alpha c [1-exp(-x^{c})]^{\alpha -1} exp(-x^{c})x^{c-1}$, for $x > 0$, $\alpha > 0$, $c >0$ \\
\hline
\end{tabular}
\begin{flushleft} Name of the statistical distributions and their PDF.
\end{flushleft}

\end{table}

\end{document}